# Which books do I like?

**Hannes Rosenbusch**[1] (h.rosenbusch@uva.nl; 0000-0002-4983-3615), **Erdem O. Meral**[1] (e.o.meral@uva.nl; 0000-0002-6326-7840)

[1]University of Amsterdam

"If I were a young person today, trying to gain a sense of myself in the world, I would do that again by reading, just as I did when I was young."
– Maya Angelou

**Finding enjoyable fiction books can be challenging, partly because stories are multi-faceted and one's own literary taste might be difficult to ascertain. Here, we introduce the ISAAC method (Introspection-Support, AI-Annotation, and Curation), a pipeline which supports fiction readers in gaining awareness of their literary preferences and finding enjoyable books. ISAAC consists of four steps: a user supplies book ratings, an AI agent researches and annotates the provided books, patterns in book enjoyment are reviewed by the user, and the AI agent recommends new books. In this proof-of-concept self-study, the authors test whether ISAAC can highlight idiosyncratic patterns in their book enjoyment, spark a deeper reflection about their literary tastes, and make accurate, personalized recommendations of enjoyable books and underexplored literary niches. Results highlight substantial advantages of ISAAC over existing methods such as an integration of automation and intuition, accurate and customizable annotations, and explainable book recommendations. Observed disadvantages are that ISAAC's outputs can elicit false self-narratives (if statistical patterns are taken at face value), that books cannot be annotated if their online documentation is lacking, and that people who are new to reading have to rely on assumed book ratings or movie ratings to power the ISAAC pipeline. We discuss additional opportunities of ISAAC-style book annotations for the study of literary trends, and the scientific classification of books and readers.**

Negative experiences with books dissuade people from reading (Hardy, 1927; Poletti et al., 2016). In order to prevent such disappointments, one has to be aware of one's own literary preferences. In this paper, we conduct a proof-of-concept self-study to determine whether an AI pipeline can improve people's awareness of their own literary preferences, and thereby help them choose more suitable books.

The question 'How do individuals evaluate books?' targets the clash of two high-dimensional systems: a person and a book. Any prediction of reading enjoyment could be based on a large number of predictors describing the person, the book, or their respective interaction terms (Bizzoni et al., 2022). Put non-statistically, readers could make sense of their own disliking of a book based on its plot ("the ending was predictable"), its writing style ("the dialogue was lacking"), their mood ("I was tired"), or a mismatch between their (self-fulfilling) narratives and the selected genre ("military sci-fi just isn't for me").

Readers might identify some issues correctly, but they might also misattribute their feelings or even misdiagnose their own general tendencies (e.g., maybe there is some military sci-fi that they would like). Further, the *relative* importance of these issues for one's ultimate reading enjoyment is, at best, difficult to ascertain (Mak et al., 2022). Thus, the high-dimensional nature of reading makes accurate introspection very challenging. Many readers appear to struggle with this issue, as it is common to give fairly vague and non-exhaustive reasons for book rejections like "being bored", "not getting into the story", or "bad writing" (Walsh & Antoniak, 2021). Even experienced editors and reviewers sometimes cite an intuitive je-ne-sais-quoi or abstract issues with the author's "voice" when explaining their own book assessments (Zhao & Wu, 2022). Some readers might be quite certain about their likes and dislikes, but that doesn't necessarily mean that their self-narratives are correct and consistently afford good book selections.

Next to the vague nature and sheer mass of variables to consider, there is another problem when trying to understand one's own literary preferences: very little training data. Unfortunately, most people don't have the resources to read many different books across many genres to ascertain their literary tastes. One can find lists of 79 or even 144 unique book genres (Emmorey, 2024; Thompson, 2019), and even if one read only a single book to evaluate an entire genre, it would require years of dedicated reading, and many disappointments, to explore them all. Considering that there are additional splitting factors between books like themes (e.g., "war", "nature", "justice") and writing styles (e.g., "lyrical", "sparse", "humorous"; Waples, 1931), it appears extremely costly to explore one's likes and dislikes, with no guarantees that one's ultimate self-narrative will result in more satisfying reading experiences.

So, how do avid readers navigate this large, blurry landscape of available books, with only their limited introspection as guidance? When people search for books they often consider books that appear to be similar to previously enjoyed books, often using the author or genre as heuristics (Mikkonen & Vakkari, 2012; Tang et al., 2014). However, within (allegedly) preferred genres, people still experience wide variations in reading enjoyment. Even within the same book series, a reader might love the debut only to loathe its sequel written by the same author and featuring the same characters and setting. Accordingly, Hollands remarked that trained librarians might carefully weigh a reader's "favorite authors, genres, and appeal factors, only to offend the reader with suggestions that violate a hidden pet peeve" (2006, p. 207).

Naturally, recommendations for people who 'don't like reading' in general, or for people that 'read everything' are challenging as well, given the complete absence of distinguishing preferences. The general helplessness of readers, when trying to match books to their (assumed) personal preferences, is apparent in the finding that many readers end up selecting books based on the cover art (Park et al., 2023; Raqi & Zainab, 2008), or switch to less challenging leisure activities (Barwise et al., 2020).

Lastly, readers might also select books that are suggested by websites like Goodreads or Amazon, where covert recommendation algorithms cluster readers and books based on overlaps in assumed readership (cf., collaborative filtering algorithms; Rana et al., 2019). While these algorithms can boost the relevance of suggested books, they are atheoretical and thus do not tell us anything about our literary tastes. Thus, they do not foster our literary self-awareness and limit our book selection to a specific website.

**What type of reader am I?**

Past research has clustered readers into latent groups like entertainers, aesthetes, and realists (cf., Mikkonen & Vakkari, 2017), which are thought to differ in their goals of immersion, artistic appreciation, and information seeking, respectively. However, reader surveys suggest that such theoretical clusters can be misleading, as there are many fundamental criteria applied by the vast majority of readers (e.g., immersion) and some additional criteria which are applied by some (e.g., intellectual stimulation). Similarly, readers simultaneously care about a book's plot, characters, setting, and writing style–and not just an individualistic subset of them (Balée, 2006; Riddell & van Dalen-Oskam, 2018).

Other attempts to divide up the readership are to consider common no-gos or must-haves (e.g., dark themes, swearing, happy endings, accessible language; Teo, 2018; Twenge et al., 2017). Some readers enjoy books that are currently widely read because public attention increases the social relevance of one's reading activities, and because knowledge about certain books comes with status benefits (Kraaykamp & Dijkstra, 1999). Other readers worry about the length of books, either stating that long books are slow/boring, or that short books don't allow for sufficient immersion (Gordon & Lu, 2021).

Literary professionals sometimes argue that the most crucial step for a fiction reader is to empathize, or at least have interest in, the main characters, as the entire story is experienced through their lens (Balée, 2006). Generally, readers find it easier to care about believable and likable protagonists (Balée, 2006; Gordon & Lu, 2021), although antagonists can also become popular characters (as they often induce agency and action into stories; Krakowiak & Oliver, 2012; Krause & Rucker, 2020).

The problem with book and character evaluations like accessibility, believability, predictability, and relatability are that they are–by definition–person dependent (Kjeldgaard-Christiansen et al., 2021). This subjectivity constitutes the key hypothesis of disposition-based theories of media enjoyment (Raney, 2013), and entails that readers need to assess the book and themselves simultaneously, accurately, and across many latent dimensions when estimating whether they will like a book, all the while trying to reduce research time and the likelihood of spoilers.

When trying to distill one's literary preferences, many believe it helpful to explicitly reflect on one's reading experiences, for instance by writing book reviews or by keeping a reading journal (Wollman-Bonilla, 1989). While this approach could, over time, improve introspection, it has two significant drawbacks: first, it demands effort, thereby potentially exacerbating people's reluctance to read. Second, a written review or journal entry is almost never a comprehensive list of relevant book characteristics, but rather a satisficing selection steered by emotion ("loved everything about this book") and confirmation biases ("yup, this genre isn't for me"). Such low-fidelity book annotations can be misleading, especially for people who only have a small sample of annotated (or remembered) books, raising the danger of false self-narratives and future disappointments.

**Study 1: Development and testing of ISAAC**
In this study, the first author attempts to transfer the task of generating high-dimensional book annotations to an AI system (cf., Jang & Jung, 2024; Yang et al., 2024). Reflecting on the outputs of this system, he will attempt to extract recurring patterns in his reading enjoyment. If this system was able to accurately *predict* his enjoyment of books, it could serve as a foundation for data-based introspection, increased awareness of literary preferences, and more successful book selections in the future.

**Method and data**
All code and data can be found here: osf.io/6euhq.

*Personal book ratings*
Over the last three years, the first author noted down the books that he read (including the ones that he didn't finish) and for each book, he tried to express his overall enjoyment on a scale from zero ("I experienced no enjoyment") to one hundred ("I experienced intense enjoyment"). Given that extreme ratings (e.g., right after giving up on a book) skewed the rating distribution (skew = -1.192), all ratings were transformed into a percentile rank before the analysis. Figure 1 describes the first author's 98 book ratings and comments.

*AI annotations*

The AI-facilitated introspection pipeline was implemented in Python code and will be referred as ISAAC (Introspection-Support, AI-Annotation, and Curation; see Figure 2). In the pipeline, each book is annotated with information collected via perplexity.ai, a search API that gathers relevant data from various sources (such as Goodreads, Wikipedia, and author websites; see results).

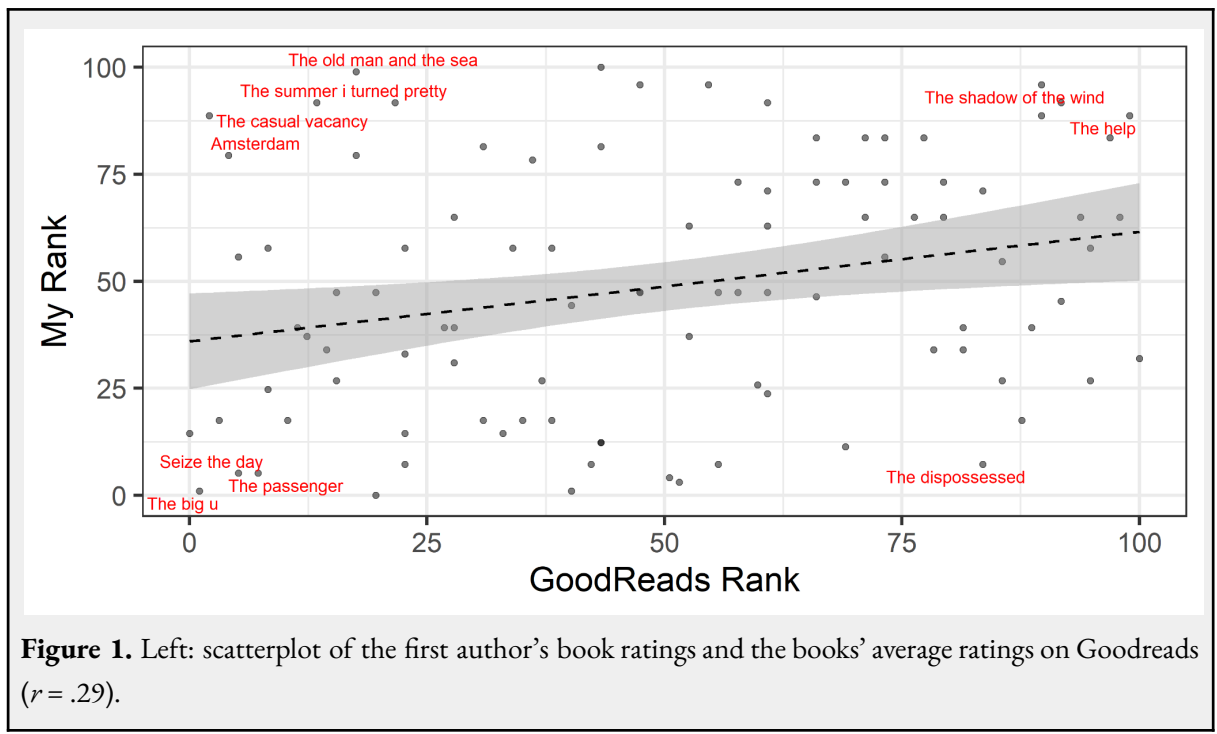

**Figure 1.** Left: scatterplot of the first author's book ratings and the books' average ratings on Goodreads ($r = .29$).

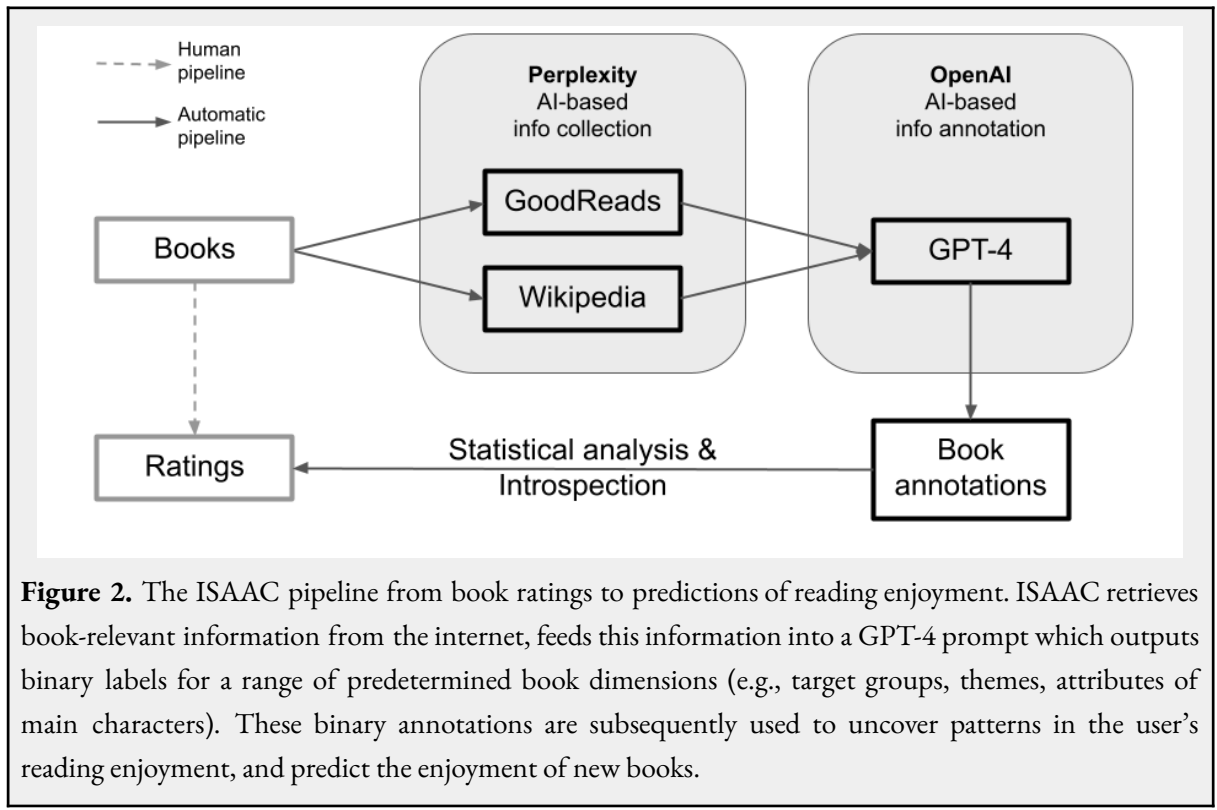

**Figure 2.** The ISAAC pipeline from book ratings to predictions of reading enjoyment. ISAAC retrieves book-relevant information from the internet, feeds this information into a GPT-4 prompt which outputs binary labels for a range of predetermined book dimensions (e.g., target groups, themes, attributes of main characters). These binary annotations are subsequently used to uncover patterns in the user's reading enjoyment, and predict the enjoyment of new books.

The set of annotations collected for each book, including the book's public metadata (e.g., genre, average review score, number of pages), and the AI-generated labels for potentially relevant book contents (e.g., themes, attributes of main character, writing style; see introduction) is shown in Table 1. AI annotations were generated based on perplexity's summary outputs, as well as the top 60 reader comments on Goodreads. In this study, ISAAC was also used to annotate the first author's private notes about each book (see last row), but these annotations were not used for predictive modeling.

*Analysis plan*

The analysis plan for Study 1 consists of four steps: 1) In order to get a standardized set of book annotations, we run each book through the ISAAC pipeline. 2) In order to verify the accuracy of the book annotations, we will manually annotate the books on a subset of dimensions and compute the agreement of AI and human-generated annotations. 3) In order to gain insights into the first author's literary preferences, we will conduct bivariate correlation analyses between annotations and book ratings (e.g., which themes were rated relatively high). 4) We will test the reliability and usefulness of these correlational findings in out-of-sample predictions (i.e., book recommendations).

**Table 1**
Annotations produced by ISAAC

| | Source | Scale |
|---|---|---|
| Average book rating | Goodreads | 1-5 star ratings |
| Number of book ratings | Goodreads | Count |
| Number pages | Goodreads | Count |
| Book genre | Goodreads | Binary |
| Top 60 Goodreads comments mention: good characters, bad characters, bad writing style, good writing style, good plot, bad plot, fast pace, slow pace, good setting, bad setting, DNF, addictive content, intellectual content | Goodreads | Proportion (AI-generated) |
| Possible target groups: women, men, romance lovers, action junkies, poetry lovers, scientists, young people, adults, social activists, history fans | Perplexity | Binary (AI-generated) |
| Style: complex, introspective, plot-focused, flowery, poetic, lots of characters, funny, experimental | Perplexity | Binary (AI-generated) |
| Mood: dark, light, happy, tragic, thrilling, serious, nostalgic, cozy, fearful, thought-provoking | Perplexity | Binary (AI-generated) |
| Main character is: teenager, adult, senior, male, female, minority member, majority member, no clear main character | Perplexity | Binary (AI-generated) |
| Themes: romance, family, war, violence, politics, prejudice, solitude, survival, magic, personal growth, women's issues, money, coming-off-age, academia | Perplexity | Binary (AI-generated) |
| Character goals: destroy an evil, establish a relationship, survive, reach political aspiration, reach professional aspiration, solve a crime, defeat an opponent, gain inner peace, understand the self, protect someone, vengeance, forgive, achieve personal growth, escape, no clear goals | Perplexity | Binary (AI-generated) |
| Character struggles against: other character, society, nature, technology, fate, supernatural, self | Perplexity | Binary (AI-generated) |
| *First author's comments: DNF, good/characters, bad/good writing style, good/bad plot, fast/slow pace, good/bad setting, addictive/intellectual content* | *First author's reading journal* | *Binary (AI-generated)* |

## Results

The results section is organized based on the four steps in the analysis plan.

### 1) Annotating the books

The ISAAC pipeline from Figure 1 ran without errors, producing a dataset of 98 rows (books) and 188 columns (annotations). Perplexity found all 98 books on Wikipedia ($N$=79), Goodreads ($N$=79), or both ($N$=65).

### 2) Agreement of human and AI annotations

The first author annotated each book on three dimensions of varying subjectivity: whether the book featured a female main character, whether the book includes historical exploration, and whether the book had a family theme. The percentage of agreement between manual and AI annotations was 96%, 89%, and 83%, respectively. A qualitative error analysis suggested that most disagreements did not stem from unequivocally false AI annotations, but rather from cases that might also produce disagreements among human annotators. For instance, all disagreements regarding the presence of a female main character occurred in books that either lacked a clear main character or featured a group of main characters (including a female character).

### 3) Predictors of book enjoyment

Figure 3 shows bivariate relationships between the annotation dimensions and reading enjoyment, as well as estimates of these relationships based on the first author's introspection.

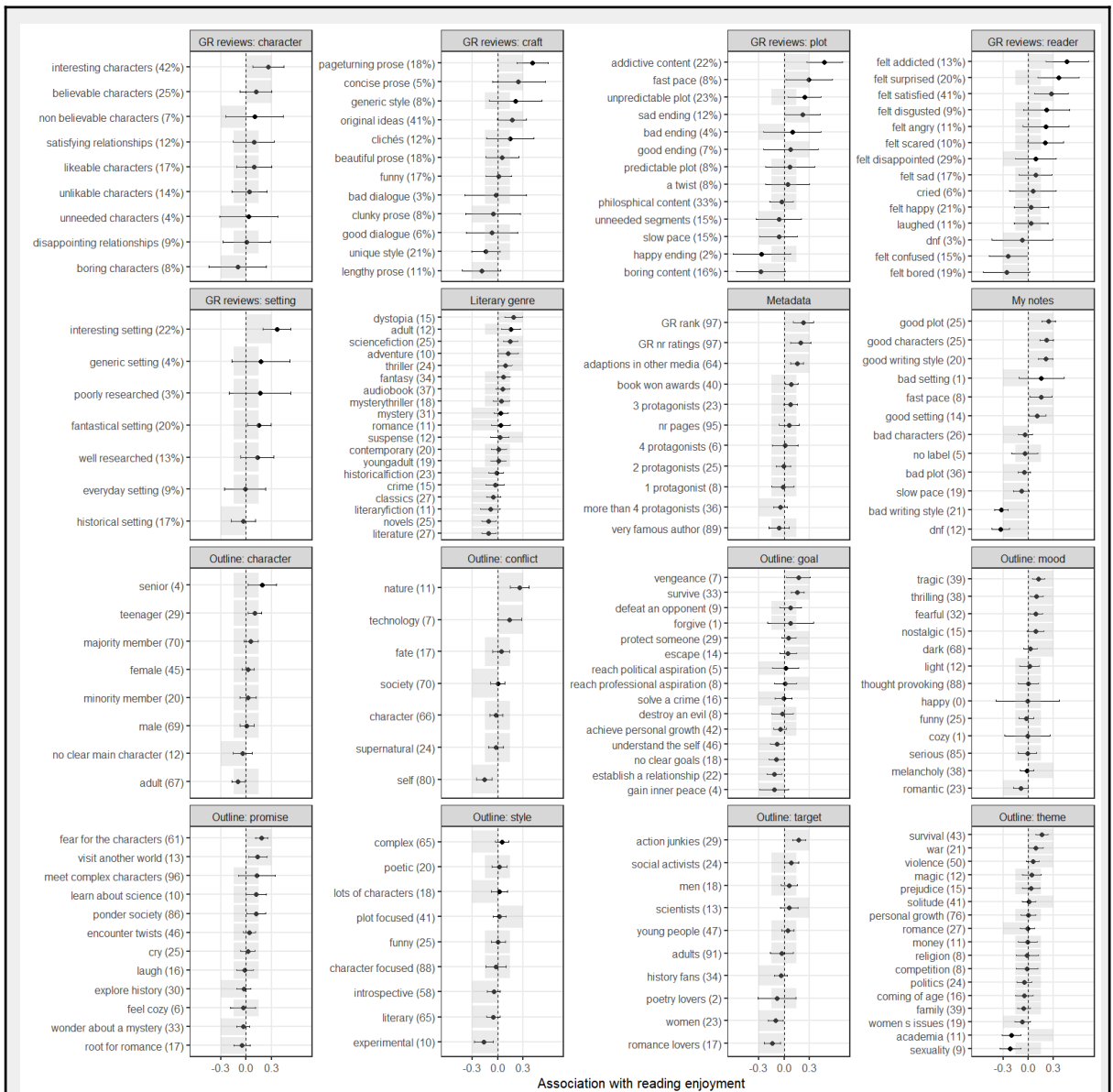

**Figure 3.** The point estimates show correlations between book annotations and reading enjoyment. Gray bands indicate how the first author judged his own literary preferences based purely on introspection. Number of books is given in parentheses. Error bars are 80% credible intervals from Bayesian simple linear regressions of reading enjoyment on annotations, with Beta(1,1) priors scaled to [-0.5, 0.5] for the focal coefficients and a standardized outcome variable.

The first author's sided expectations based on introspection coincided with the statistics about his reading enjoyment in 88% of cases (if measured as same-side deviation from a zero correlation).

*4) Prediction of book enjoyment*
Given the small sample and effect sizes, ISAAC's bivariate analyses need to be interpreted with some caution. One way to assess their overall reliability and usefulness is to test whether they can accurately predict the enjoyment of new books. In the realm of machine learning, this requires fitting a statistical model on one dataset, and measuring the predictive accuracy of the model on a new data set.

Here, we fit a statistical prediction model repeatedly, each time leaving out one book from the first author's book collection. The enjoyability of this left-out book is subsequently predicted and the accuracy is noted (cf., leave-one-out cross-validation).

Figure 4 shows the model performance of three models: a ridge regression, a random forest model, and a simple linear regression that predicts book enjoyment with only the average Goodreads rating. Model hyperparameters were selected based on nested cross-validation.

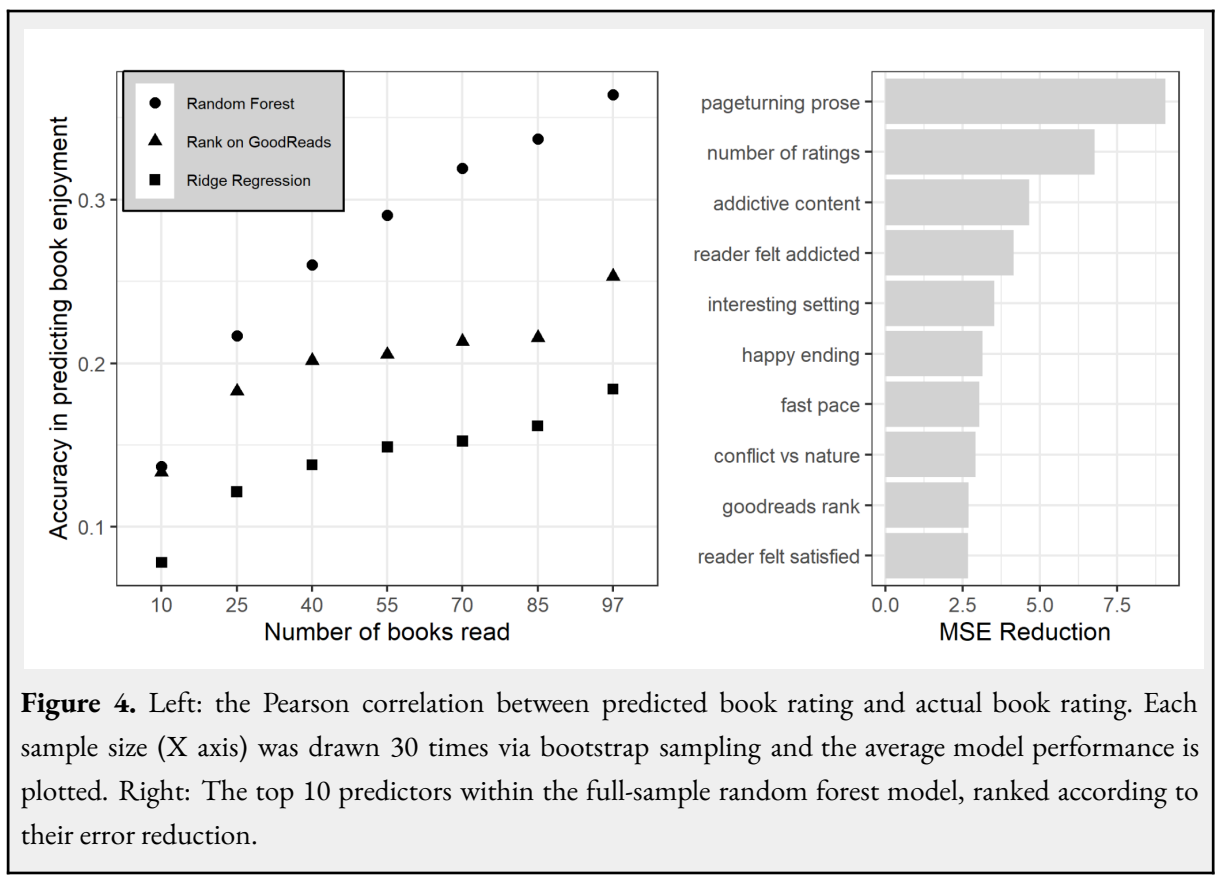

**Figure 4.** Left: the Pearson correlation between predicted book rating and actual book rating. Each sample size (X axis) was drawn 30 times via bootstrap sampling and the average model performance is plotted. Right: The top 10 predictors within the full-sample random forest model, ranked according to their error reduction.

**Interim discussion**

The insights from the first ISAAC application are that automatic book annotations are possible if the book's attributes and contents are documented online. Using current state-of-the-art LLMs, such annotations are fairly accurate, meaning the agreement rate with a human annotator appears close to between-human agreement.

The associations between specific book attributes and the first author's reading enjoyment appear to be mostly small. Given that large effects would require strong homogeneity within annotations and stark differences between annotations (e.g., loving every book about nature and hating all other books), effects will likely be small for most readers. Further, the uncertainty bars are often quite large because the first author, like most readers, has only read a few books with certain characteristics (e.g., four books with a senior main character and two books for poetry lovers). Thus, while the first author might be quick to form an opinion about certain book genres, his book ratings highlight that evidence accumulation for one's literary preferences is a slow process. Notable deviations between the data-based and introspection-based preferences were that books with an academia theme ranked lower than expected. Conversely, the first author had assumed that books about crime, romance, and historical fiction would rank relatively low, but they turned out to be centered around the average. Further, he did not expect that the prevalence of 'surprise' among Goodreads readers would be associated with his book enjoyment, but the evidence suggests that readers' reports of surprise and their difficulties to predict the plot were positive predictors of enjoyment. Diving deeper into the formation of the first author's previous beliefs, it appears that a few salient books distorted his perception of certain genres. For instance, it is possible that certain moody, campus novels made him overestimate his enjoyment of novels with academic themes, whereas a few clichéd romance and crime stories led him to judge these genres too harshly and overlook books in the genres that he had actually enjoyed.

Regarding book recommendations, ISAAC proved useful for predicting the enjoyment of new books. The achieved accuracy surpassed common heuristics (e.g., ranking books based on genre or average review score). Further, rating more books appears to improve the forecasting model even beyond the first one hundred books–a finding that might differ for other readers with more (or less) diversity in their books.

**Study 2: Replication on different reader**

As reading enjoyment, and the act of introspection, could be fairly unique for the first author's case (given his work as a literature researcher and writer), the ISAAC pipeline was applied to another reader–the second author–who joined the project after the completion of the analyses above. The second author first listed his assumed literary preferences (see pre-registration in supplementary materials; see Figure 5), and then provided his 188 book ratings of English books accumulated on the Goodreads website over the last ten years. The second author added five book attributes to the annotation dimensions that he assumed to correlate with his reading enjoyment. Otherwise, the ISAAC pipeline was applied as described above.

**Results**

The scraping and annotation step within ISAAC produced a dataset of 188 rows (books) and 181 columns (annotations). Perplexity found 95% of the books on either Wikipedia ($N$=136), Goodreads ($N$=137), or both ($N$=95). Information about the other books was limited to author websites, commercial websites, blogs, and reviews. Figure 5 shows bivariate relationships between annotations and book ratings, as well as the second author's pre-registered estimates based purely on introspection.

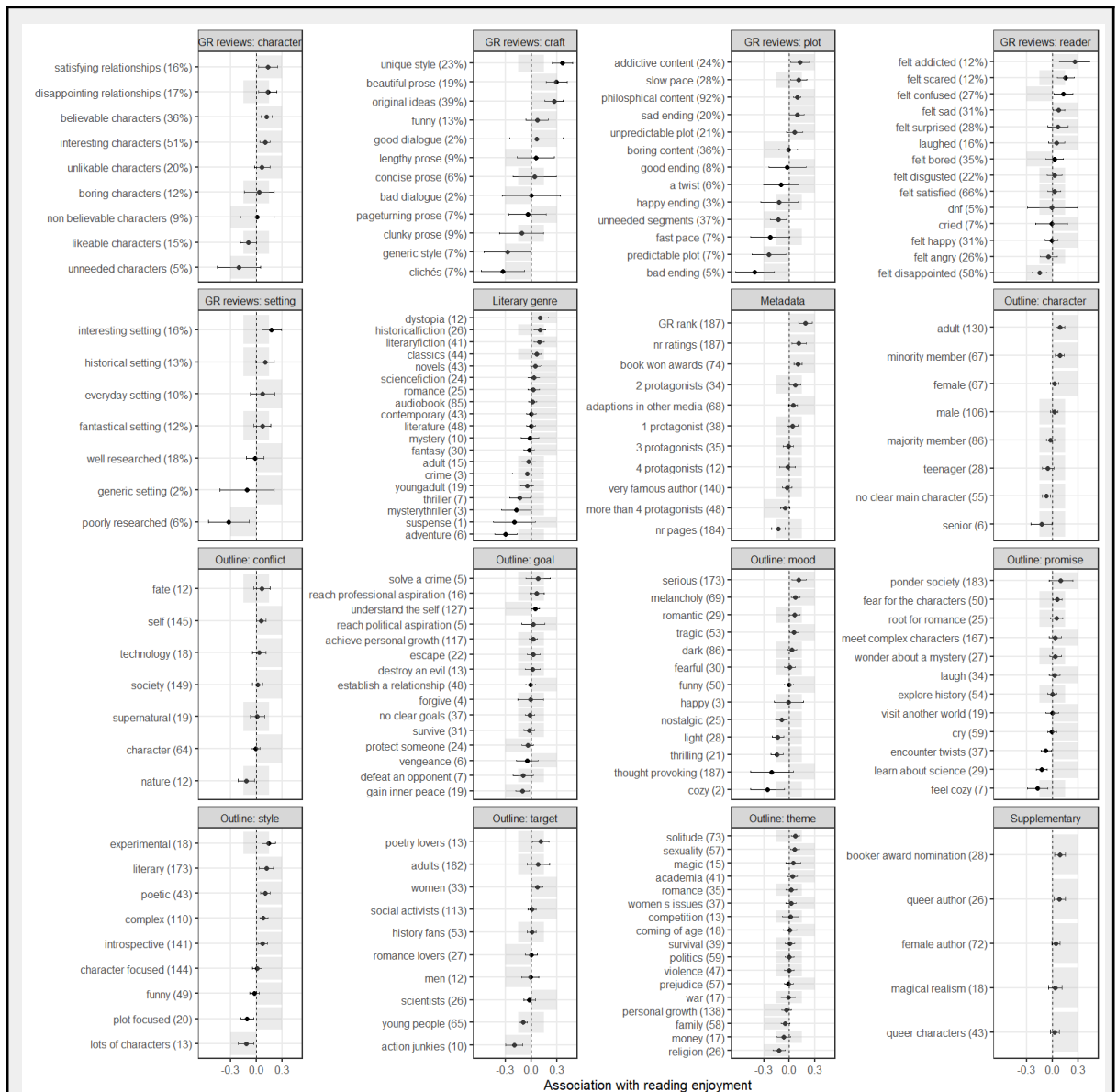

**Figure 5.** Replication of Figure 4 using the data from the second author. Note that the second author pre-registered expectations about five book dimensions that weren't included in the first annotation set (see bottom right panel).

The second author's sided expectations coincide with the data 77% of the time (if measured as same-side deviation from a zero correlation). While some covariates are shared with the first author (e.g., positive correlation of enjoyment with public review scores, addictiveness, and dark dystopian stories), others deviate (e.g., more appreciation of form and literary fiction). One surprise noted by the second reader were the positive effects of unique, experimental writing styles. The second author distinctly remembers having difficulty engaging with some experimental books. An explanation might be that he didn't finish said book and then, unlike the first author, didn't rate them. Thus, the appreciation of "unique", "experimental", "confusing" books might be inflated through a survivorship bias as such books needed to be especially enjoyable in order to be finished and rated (Figure 5, panel on GR Reviews: Reader).

Furthermore, in recent years the second author focused on reading books written by queer authors or include queer (main) characters, and books that either won or were nominated for a Booker prize (a theme of his book club). These informed his expectations in the supplementary panel in Figure 5. The main surprise here was the small size of the effects. One reason for this could be that the second author is more critical towards this literature, which may explain why the anticipated positive effects are small or nonexistent.

For out-of-sample predictions of book enjoyment, we again tested three models and varying sample sizes (see Figure 6).

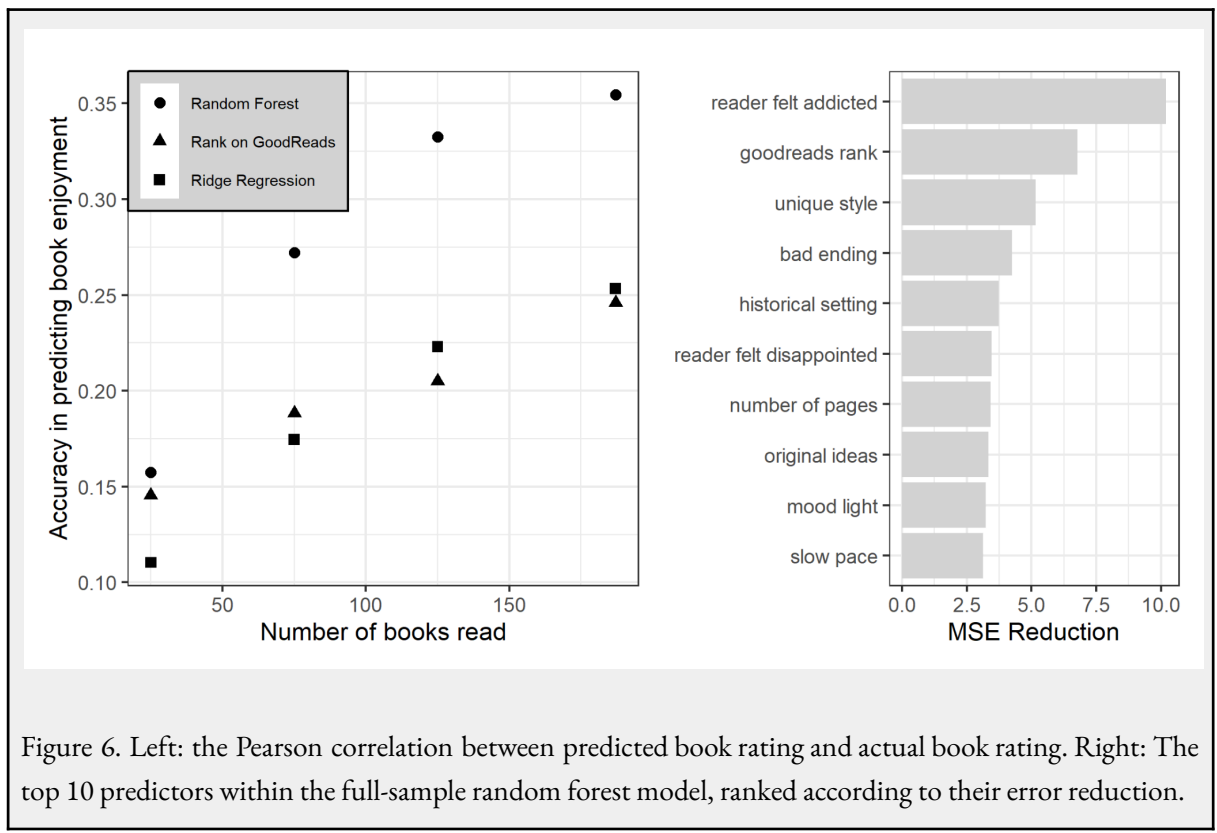

Figure 6. Left: the Pearson correlation between predicted book rating and actual book rating. Right: The top 10 predictors within the full-sample random forest model, ranked according to their error reduction.

## General discussion

Identifying the types of books one enjoys is a difficult and lengthy process including many disappointing reading experiences swaying readers towards other forms of entertainment. Known methods to improve introspection and awareness of one's literary preferences are to keep notes or a reading journal. Unfortunately, this requires substantial time and effort. Alternatively, one could rely on a black-box recommendation algorithm to serve up books that are similar to one's previous purchases. However, this method does not enhance the reader's self-awareness and thus binds them to the platform and its recommendation engine.

The ISAAC method, presented in this paper, integrates automation and introspection by collecting information about rated books and annotating them on customizable dimensions (e.g., "this book had a female protagonist" or "this book is a story about war"). The analysis section of the ISAAC pipeline highlights patterns among the books that the reader previously enjoyed (or disliked) thereby supporting introspection. Lastly,

ISAAC predicts the enjoyment of unread books in a relatively transparent and explainable way.

Our first two applications of the new method resulted in multiple insights. Firstly, the annotation of the books works well, provided that the requested book attributes are well-documented online. Annotation dimensions that are vague or otherwise challenging ("Would Michelle Obama recommend this book?") can be expected to be fairly noisy, regardless whether annotators are humans or AIs.

Second, both authors were re-affirmed in some of their beliefs about their reading preferences but, crucially, also found unexpected patterns that warrant (re-)consideration of specific themes and genres. ISAAC allowed us to test hypotheses about our tastes with custom annotations, while large standard errors drew our attention to themes and genres that warrant increased exploration. Importantly, it also became obvious that statistical patterns should not replace one's self-narrative. For most readers, small sample sizes will preclude strong conclusions and some crucial data will not be considered by ISAAC (e.g., a reader might avoid books about war because of real-life experiences). Rather, ISAAC's outputs invite reflection about how certain self-narratives might have come about (e.g., "What were my reasons for rejecting crime stories?").

Thirdly, the prediction accuracy for new books surpassed common heuristics (e.g., ranking books based on genre or average review score) meaning that the ISAAC pipeline could function as a viable recommendation system. Even after a hundred books, the prediction accuracy still increased. The associated disadvantage is that new readers will benefit less from the ISAAC method than readers who have rated many books. There are ways to bridge this "cold start" problem, for instance by constraining the prediction model with one's prior assumptions or, more simply, by giving "hypothetical" ratings to books that the user hasn't actually read yet but still has a preliminary opinion about (e.g., based on summaries, popular culture, similar books; cf., Koolen et al., 2020). Further, it is of note that a recommendation system does not need (and rarely has) perfect prediction accuracy. Rather, it quickly ranks a huge number of books of which only the very top prediction is used as a recommendation. The ability to correctly rank books of average enjoyability is usually unimportant.

A more severe requirement of the ISAAC pipeline is the quality of the input data. It is important to provide representative samples of book ratings as an a-priori selection (e.g., not only books that were finished/enjoyed; cf., results for the second author). Similarly, low-fidelity book ratings (e.g., given without reading the book) can do more harm than good, if the reader mispredicts their own book enjoyment.

Future extensions of the pipeline are to replicate the analyses with a large number of readers willing to donate their book ratings for scientific inquiry and engage in introspective interaction with ISAAC. This would allow us to optimize the user introspection step (e.g., through result-based prompts), document different model learning curves and prediction accuracies. If it was possible to devise a default selection of annotation dimensions that are predictive *across* readers, this would enable practitioners to build a database of annotated books which would in turn enable fast and cheap book recommendations, even with person-specific prediction models. However, silver-bullet annotations that correlate very strongly with book enjoyment (even just for some users) are likely rare, with the exception of readers with very strong aversions against certain themes and contents.

Further, new methods of overcoming the "cold start" problem can be explored. When introspecting about our literary preferences, we do not *only* rely on past reading experiences. Our general interests and personality traits steer our enjoyment of certain stories across media formats (Rentfrow et al., 2011). It would be relatively straightforward to append ratings of movies and TV shows to one's book ratings as they often share the same annotation dimensions (e.g., genres, themes, tone, character attributes). Thus, they could be used without any alterations to the ISAAC pipeline for analyses and predictions of the types of *stories* that one enjoys (rather than only books; Li et al., 2009).

Lastly, and most simply, future research and applications of the pipeline will require a user-friendly frontend that allows for easy upload of one's book ratings (e.g., Goodreads exports) and that facilitate the data-based introspection through visual integration of expected and data-based preferences. Additional features could be the highlighting of radical mispredictions (i.e., surprising likes/dislikes) as well as corpus-based book recommendations. It is also worthwhile giving users the option to delete annotation dimensions in between the introspection and recommendation steps, as they might, for instance, mistrust their predictive validity. It is also possible to alter the prediction model to output the most informative, as opposed to the most enjoyable, book (e.g., ranked by maximal uncertainty reduction across annotations). This could address fiction readers' desire to optimally explore (rather than exploit) their literary preferences (Murdock et al., 2017).

AI annotations in the style of ISAAC, can be used for adjacent efforts in computational literary studies. They could, for instance, allow for renewed attempts of typifying books and readers (Birdi & Ford, 2018; Miesen, 2004). Similarly, the study of historical trends can be facilitated through fast and accurate book annotations (e.g., changes in genre definitions, King, 2021; correlates of book contents and reviews with author or audience characteristics, Thelwall, 2019).

## Conclusion

In order to read, one has to find enjoyable books, and in order to know what's enjoyable, one has to be aware of one's literary preferences. We devised an AI supported system that highlights a reader's likes and dislikes based on previous book ratings and contrasts them with the reader's assumed preferences, thereby facilitating improved introspection. The system has numerous advantages like relatively low requirements of time and effort, customizability, transparency, integration with human intuition, and highlighting of underexplored literary niches. Users can extend their input data with ratings of other media types (e.g., movies) and recommendations are based on covert, automatic internet searches thereby avoiding spoilers. Disadvantages of the system are that it can steer, but not avoid, a user's manual exploration of the literary landscape, and that rated books must be sufficiently documented on the internet.

## References

Balée, S. (2006). *Textual Pleasures and Pet Peeves*. JSTOR. https://www.jstor.org/stable/20464508

Barwise, P., Bellman, S., & Beal, V. (2020). Why do people watch so much television and video?: implications for the future of viewing and advertising. *Journal of Advertising Research, 60*(2), 121-134.

Birdi, B., & Ford, N. (2018). Towards a new sociological model of fiction reading. *Journal of the Association for Information Science and Technology, 69*(11), 1291-1303.

Bizzoni, Y., Lassen, I. M. S., Peura, T., Thomsen, M., & Nielbo, K. (2022). Predicting Literary Quality How Perspectivist Should We Be? *NLPERSPECTIVES*.

Emmorey, B. R. (2024, August 7). *Book genres: A complete guide to 79+ fiction and nonfiction genres*. Self-Publishing School. https://self-publishingschool.com/book-genres/

Gordon, C., & Lu, Y.-L. (2021). "I Hate to Read-Or Do I?" Low-Achievers and Their Reading. *IASL Annual Conference Proceedings*. https://journals.library.ualberta.ca/slw/index.php/iasl/article/view/7972

Hardy, M. (1927). Right attitude toward books and taste in reading in the primary school. *The Elementary School Journal, 27*(10), 745-750.

Hollands, N. (2006). Improving the model for interactive readers' advisory service. *Reference & User Services Quarterly, 45*(3), 205-212.

Jang, W., & Jung, S. (2024, November). Evaluating LLM Performance in Character Analysis: A Study of Artificial Beings in Recent Korean Science Fiction. In *Proceedings of the 4th International Conference on Natural Language Processing for Digital Humanities* (pp. 339-351).

King, R. S. (2021). The Scale of Genre. *New Literary History, 52*(2), 261-284.

Kjeldgaard-Christiansen, J., Fiskaali, A., Høgh-Olesen, H., Johnson, J. A., Smith, M., & Clasen, M. (2021). Do dark personalities prefer dark characters? A personality psychological approach to positive engagement with fictional villainy. *Poetics, 85*, 101511.

Koolen, C., van Dalen-Oskam, K., van Cranenburgh, A., & Nagelhout, E. (2020).

Literary quality in the eye of the Dutch reader: The National Reader Survey. *Poetics, 79*, 101439.

Krakowiak, K. M., & Oliver, M. B. (2012). When good characters do bad things: Examining the effect of moral ambiguity on enjoyment. *Journal of Communication, 62*(1), 117-135.

Krause, R. J., & Rucker, D. D. (2020). Can bad be good? The attraction of a darker self. *Psychological Science, 31*(5), 518-530.

Li, B., Yang, Q., & Xue, X. (2009, June). Can movies and books collaborate? cross-domain collaborative filtering for sparsity reduction. In *Twenty-First international joint conference on artificial intelligence.*

Mak, M., Faber, M., & Willems, R. M. (2022). Different routes to liking: how readers arrive at narrative evaluations. *Cognitive research: principles and implications, 7*(1), 72.

Miesen, H. (2004). Fiction readers' appreciation of text attributes in literary and popular novels: Some empirical findings. *International Journal of Arts Management*, 45-56.

Mikkonen, A., & Vakkari, P. (2012, August). Readers' search strategies for accessing books in public libraries. In *Proceedings of the 4th information interaction in context symposium* (pp. 214-223).

Mikkonen, A., & Vakkari, P. (2017). Reader characteristics, behavior, and success in fiction book search. *Journal of the Association for Information Science and Technology, 68*(9), 2154-2165.

Murdock, J., Allen, C., & DeDeo, S. (2017). Exploration and exploitation of Victorian science in Darwin's reading notebooks. *Cognition, 159*, 117-126.

Poletti, A., Seaboyer, J., Kennedy, R., Barnett, T., & Douglas, K. (2016). The affects of not reading: Hating characters, being bored, feeling stupid. *Arts and Humanities in Higher Education*, *15*(2), 231–247. https://doi.org/10.1177/1474022214556898

Rana, A., & Deeba, K. (2019, November). Online book recommendation system using collaborative filtering (with Jaccard similarity). In *Journal of Physics: Conference Series* (Vol. 1362, No. 1, p. 012130). IOP Publishing.

Raney, A. A. (2013). The psychology of disposition-based theories of media enjoyment. In *Psychology of Entertainment* (pp. 137-150). Routledge.

Raqi, S. A., & Zainab, A. N. (2008). Observing Strategies Used by Children When Selecting Books to Browse, Read or Borrow. *Journal of Educational Media & Library Sciences, 45(4).*

Rentfrow, P. J., Goldberg, L. R., & Zilca, R. (2011). Listening, Watching, and Reading: The Structure and Correlates of Entertainment Preferences. *Journal of Personality, 79*(2), 223–258. https://doi.org/10.1111/j.1467-6494.2010.00662.x

Riddell, A., & van Dalen-Oskam, K. (2018). Readers and their roles: Evidence from readers of contemporary fiction in the Netherlands. *PloS one, 13*(7), e0201157.

Thelwall, M. (2019). Reader and author gender and genre in Goodreads. Journal of *Librarianship and Information Science, 51*(2), 403-430.

Thompson, T. (2019, February 10). *Book genre encyclopedia: 144 genres and subgenres explained*. ServiceScape. https://www.servicescape.com/blog/book-genre-encyclopedia-144-genres-and-subgenres-explained

Teo, H. M. (2018). The Contemporary Anglophone Romance Genre. In Oxford *Research Encyclopedia of Literature*.

Twenge, J. M., VanLandingham, H., & Keith Campbell, W. (2017). The seven words you can never say on television: Increases in the use of swear words in American books, 1950-2008. *Sage Open, 7*(3), 2158244017723689.

Wollman-Bonilla, J. E. (1989). Reading journals: Invitations to participate in literature. *The Reading Teacher, 43*(2), 112-120.

Yang, Z., Liu, Z., Zhang, J., Lu, C., Tai, J.,
 Zhong, T., ... & Liu, T. (2024). Analyzing
 nobel prize literature with large
 language models. *arXiv preprint
 arXiv:2410.18142*.

Zhao, C. G., & Wu, J. (2022). Perceptions of
 authorial voice: Why discrepancies exist.
 *Assessing Writing, 53*, 100632.